\documentclass[10pt]{article}
\usepackage{latexsym,graphicx}
\usepackage{latexsym,graphicx}
\newcommand{\be}{\begin{equation}}
\newcommand{\ee}{\end{equation}}
\def\n{\noindent}
\catcode `\@=11 \catcode `\@=12
\begin{document}
\begin{center}
\large{\bf {A theoretically-explained new-variant of Modified-Newtonian-Dynamics (MOND)}}\\
\vspace{10mm}
\normalsize{R. C. Gupta $^1$ and Anirudh Pradhan $^2$} \\
\vspace{5mm} \normalsize{$^1$ GLA Institute of Technology and Management (GLAITM), \\
Mathura-281 406, India} \\
\normalsize{E-mail: rcg\_iet@hotmail.com, rcgupta@glaitm.org}\\
\vspace{5mm} \normalsize{$^2$ Department of Mathematics, Hindu
P. G. College,\\Zamania-232 331, Ghazipur, India} \\
\normalsize{E-mail: acpradhan@yahoo.com, apradhan@imsc.res.in}\\
\end{center}
\vspace{10mm}
\begin{abstract}
It is surprising that we hardly know only $5\%$ of the universe.
Rest of the universe is made up of  $70\%$ of dark-energy and $25\%$
of dark-matter. Dark-energy is responsible for acceleration of the
expanding universe; whereas dark-matter is said to be necessary as
extra-mass of bizarre-properties to explain the anomalous
rotational-velocity of galaxy. Though the existence of dark-energy
has gradually been accepted in scientific community, but the
candidates for dark-matter have not been found as yet  and are too
crazy to be accepted. Thus, it is obvious to look for an alternative
theory  in place of dark-matter. Israel-scientist M. Milgrom has
suggested a `Modified Newtonian Dynamics (MOND)' which appears to be
highly successful for explaining the anomalous rotational-velocity.
But unfortunately MOND lacks theoretical support. The MOND, in-fact,
is (empirical) modification of Newtonian-Dynamics through
modification in the kinematical acceleration term `a' (which is
normally taken as $a=\frac{v^{2}}{r}$) as effective kinematic
acceleration  $a_{effective} = a \mu (\frac{a}{a_{0}})$ , wherein
the $\mu$-function is 1 for usual-values of accelerations but equals
to $\frac{a}{a_{0}} (\ll 1)$ if the acceleration `a' is extremely-low
lower than a critical value $a_0 (10^{-10} m/s^{2})$. In the present
paper, a novel variant of MOND is proposed with theoretical backing;
wherein with the consideration of universe's  acceleration $a_{d}$
due to dark-energy, a new type of $\mu$-function on
theoretical-basis emerges out leading to $a_{effective} = a\left(1 -
K \frac{a_{0}}{a}\right)$. The proposed theoretical-MOND model too is able to
fairly explain qualitatively the more-or-less `flat' velocity-curve
of galaxy-rotation, and is also able to predict a dip (minimum) on
the curve.
\end{abstract}
\smallskip
\n Key words : Cosmology, Dark Matter, MOND, Dark Energy \\
\n PACS: 98.80.-k, 95.36.+x, 95.35.+d \\
\section{Introduction}
How much matter and energy are there in the universe ? It is now well established that
universe-expansion began with a big-bang \cite{ref1}. The ultimate fate of the universe
depends on the universe's matter \& energy density ($\rho$) as compared to a certain value
called critical-density ($\rho_{c}$). If $\rho > \rho_{c}$, the universe is said to be `closed';
and its expansion will slow down (decelerate) and start contracting leading finally to a
big-crunch (meaning  hot-death of the universe). If $\rho < \rho_{c}$, the universe is said
to be `open'; and will expand forever even much faster (leading the universe to cold-death).
If $\rho = \rho_{c}$, the universe is said to be `flat'; and will continue to expand but not
that-fast to lead to cold-death soon. The ratio $k = \frac{\rho}{\rho_{c}}$, determine that
whether the universe is closed ($k>1$), open ($k<1$) or flat ($k=1$). It has been estimated
that the universe would have collapsed (to hot-death) much sooner than the present-age of
the universe if $ k>1$; and it would have cooled down (to cold-death) much earlier than the
present-age of the universe if $k < 1$. The present-age (14 billion years) constraint of the
universe, compel the scientists to believe that $k = 1$, i.e., the universe must be
flat \cite{ref1}. \\\\
Once agreed-upon that the universe density $\rho = \rho_{c}$, the
next question arises that `what is the universe made of'? Estimation
of visible-type matter like galaxies, stars, planets etc. hardly
leads only to about $2\%$ of $\rho_{c}$; and when other all such
things like inter-galactic gases, black-hole, white-dwarf,
neutron-stars etc. are also included, the estimate hardly reaches a
mere $5\%$ of $\rho_{c}$. What is then $95\%$ of the remaining-part
? It seems invisible and unknown, hence thought as dark
constituent(s). Scientists have, presently, estimated that the
major-chunk of the universe \cite{ref2,ref3} is repulsive-gravity
type dark-energy (about $70\%$) causing the universe's  accelerated
expansion, and the rest is non-baryonic invisible but
gravitating dark-matter (about $25\%$) causing anomalous high rotational-speed of galaxies.\\\\
Recognition of dark-energy mainly through Supernovae (SNe Ia)
observations [4-9], galaxy cluster measurements \cite{ref10} and
cosmic microwave background (CMB and WMAP) data \cite{ref11,ref12}
is comparatively a recent affair; the possibility (necessity) of
dark-matter was anticipated quite-early (in 1935 by Fritz Zwicky),
but the work of Fritz Zwicky \cite{ref13,ref14} was largely ignored.
Much later (1960s to 1980s), it is the female-scientist Veera Rubins
concentrated efforts \cite{ref15,ref16} which made the
male-scientists to finally take the possibility of dark-matter
seriously. Now, based on the rotational-velocity curves of galaxies
and galaxy-clusters, the scientific community generally believe the
necessity of dark-matter. But the next question is rather yet
unanswered that `what makes the dark matter'?  Dark matter seems to
be non-baryonic. Several schools of thought have emerged
\cite{ref3}, none of these very satisfactory; can be grouped in two
categories: (i) Hot dark matter and (ii) Cold dark matter. Neutrino
is the main candidate for hot dark matter, but unable to take-up the
full account. Exotic sub-particles, the candidates for the cold dark
matter are: `weakly interactive massive particles' (WIMPs) and
`massive astrophysical compact halo objects' (MACHOs); WIMPs include
exotic extremely-long `axion' and new-breed of particles named
`s-particles'(super-partner of particles, a possibility
based on recent  super-symmetry theory).\\\\
Though there seems a need of dark-matter inside the galaxy to
explain the anomaly of galaxy's rotational-speed (higher
rotational-speed requires more mass inside), but the candidates of
dark-mater are so strange to be believed (as if, dark-matter is
castle-in-air). We wish if there is any way out, to explain the
anomaly of  rotational-speed of galaxies, without the need of
dark-matter.  In fact {\it that is what}  the Israel-scientist
Mordehai Milgrom did and proposed a `Modified Newtonian Dynamics
(MOND)'. We are now left with two alternatives/options: (1) Believe
in the existence of the crazy dark-matter, and try to find out what
(candidate) makes the dark-matter, OR (2) Use Milgrom's MOND theory
(or its variant)  which eliminates the need
of dark-matter altogether.\\\\
There would appear some confusions about the `acceleration symbols
and term(s)' and problem in the understanding the Milgrom's MOND
theory (section-3) and  its new variant proposed in this paper
(section-4), unless the term acceleration' is re-examined,
re-defined and clarified as follows (in section-2).
\section{Acceleration Re-defined}
It is said that {\it {`everything thing on earth and beyond, is governed by law of mechanics'}}.
Mechanics is subdivided into `Statics' and `Dynamics'. Dynamics is categorized further  in two
categories: `Kinematics' and `Kinetics'. The `kinematics' is that part of dynamics wherein mass
is of no concern, such as displacement and velocity; whereas `kinetic' refers to the dynamics
which depend on mass, such as force and moment-of-inertia. It has not been appreciated earlier
in physics that `acceleration', in fact belongs to both i.e., there are two types of acceleration
viz., `kinematic acceleration' and `kinetic acceleration'. If a particle is moving in circular-path,
then due to kinematical change in velocity towards center, it would have a `kinematic-acceleration $a$'
towards center commonly referred as centripetal-acceleration  $\frac{v^{2}}{r}$, which is
also sometimes referred as pseudo centrifugal-acceleration. On the other hand the gravitational-
acceleration, say on a star in spiral galaxy-arm, can be found using Newton's second law as
`force divided by star-mass $m$' as $a_{n}= \frac{GM}{r^{2}}$, which is in fact  `kinetic-
acceleration $a_{n}$' (the subscript n in $a_{n}$ refers to N of  Newton's law). Normally, both
of these `kinetic' and `kinematic' accelerations are equal hence considered as same, that is why
there has been no need for any differentiation. But, time has come to differentiate between
`kinetic' and `kinematic' accelerations, for better understanding of Milgrom's MOND and herewith
proposed MOND-variant  wherein there would be some changes in the kinematic-acceleration; the
new effective kinematic-acceleration $\bf a_{effective} = a ~ (\mu ~ function)$. The $\mu$-function
is taken to be $1$  for usually encountered acceleration, but less than $1$  if the acceleration
is extremely small. In all the situations, the kinetic-acceleration equals the
effective-kinematic-acceleration, mathematically speaking, $\bf a_{n} = a_{effective}$; the right-hand-side
will be equal to kinematic acceleration `$a$' or less than `$a$' depending whether the value of
usual kinematic acceleration `$a$' is greater than a critical value ($a_{0}$) or less than that
(See section-3 of MOND wherein it is described in detail).

\begin{figure}[ht]
\centering
\includegraphics[width=12cm,height=10cm,angle=0]{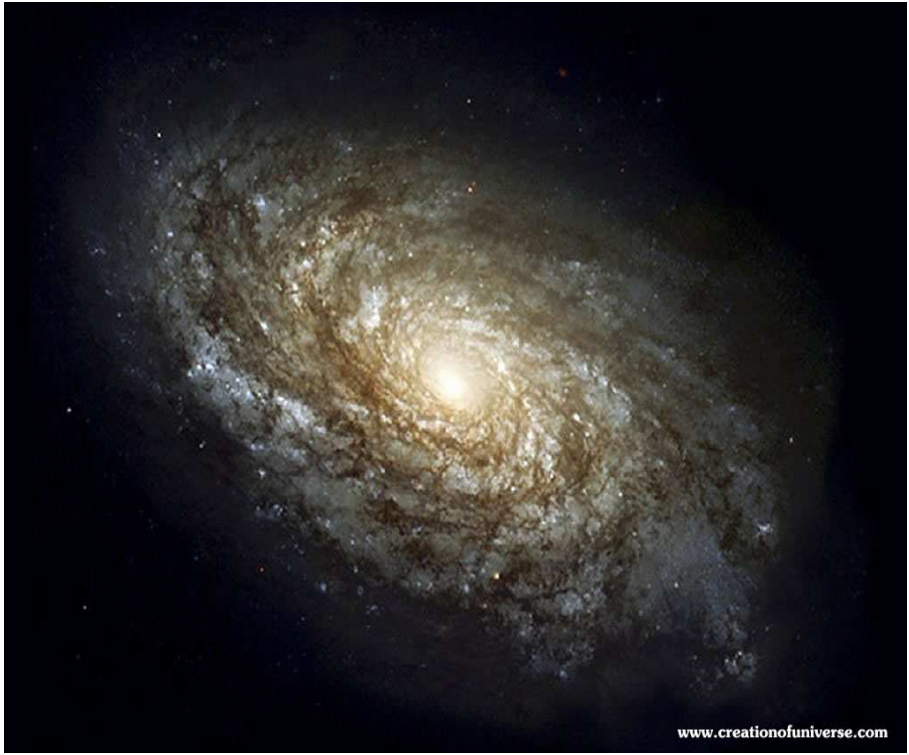} \\
\caption{A Typical Spiral Galaxy}
\end{figure}
\section{Modified Newtonian Dynamics (MOND)}
\underline{Mo}dified \underline{N}ewtonian \underline{D}ynamics (briefly abbreviated as MOND) was
proposed by Milgrom \cite{ref17,ref18} as modification in `kinematic acceleration', at extremely
low acceleration, to explain the galaxy-rotation problem. MOND eliminates the need of dark-matter. \\\\
Detailed observations of rotational speed of galaxies (Fig.1),
\begin{figure}[ht]
\centering
\includegraphics[width=11cm,height=10cm,angle=0]{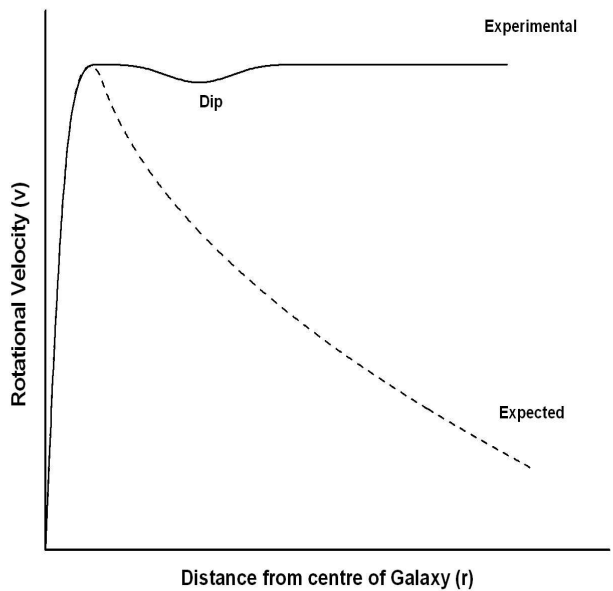}
\caption{Rotational velocity versus distance from Galaxy-centre:
expected curve based on Newtonian dynamics and experimental `flat'
curve with a dip}
\end{figure}
in 1980, made it clear that galaxies do not exhibit the same pattern
of decreasing orbital velocity  with increasing distance from the
center-of-mass as observed in the Solar-system. A spiral galaxy
(Fig.1) consists of a bulge of stars at the center with a vast
number of stars orbiting around the central solid-type big-lump. If
the orbits of the galaxy's stars were solely governed by the
central-gravitational force, it was expected that the stars at the
outer edge of the galaxy would have a much lower orbital-velocity
than  that of those near the middle of the galaxy. In the observed
galaxies, this velocity pattern is not noticed; stars near the outer
edge were found to be orbiting  at about the same-speed as the inner
stars near the middle (Fig.2). This ``flattening of galaxy's
rotational-curve' requires invisible dark-matter within the galaxy,
or necessitates the use of Modified Newtonian Dynamics (MOND). \\\\
The MOND theory, in fact, modifies the `kinematic-acceleration' with
a $\mu$-function as follows, specially when acceleration is
extremely small $a \ll a_{0}$ ($a_{0}\approx10^{-10} m/s^{2}$).
According to MOND, the modified `effective kinematic-acceleration'
is given as
\begin{equation}
\label{eq1} a_{effective} = a ~ \mu{\left(\frac{a}{a_{0}}\right)},
\end{equation}
$\mu{(\frac{a}{a_{0}})}$ is termed as $\mu$-function and $a_{0}
\approx 10^{-10} m/s^{2}$. The function $\mu (\frac{a}{a_{0}})$
turns out to be as follows.
\begin{eqnarray}
 \mu ~ \left(\frac{a}{a_{0}}\right) &=& 1 \qquad \textrm {for $a \gg a_{0}$}, \nonumber \\
&=& \frac{a}{a_{0}} \qquad \textrm {for $a \ll a_{0}$}.
\end{eqnarray}
Now applying MOND to the gravitational attraction-force between a star to the central galaxy core
(of mass M),
\[
 a_{n} = a_{effective}
\]
i.e.
 \begin{equation}
\label{eq3} \frac{GM}{r^{2}} = a ~ \mu {\left(\frac{a}{a_{0}}\right)}
\end{equation}
At large distance `r' at the galaxy outskirt, the kinematical
acceleration `a' is extremely-small smaller than $10^{-10} m/s^{2}$,
i.e., $a \ll a_{0}$, hence the function
$\mu{\left(\frac{a}{a_{0}}\right)} = \frac{a}{a_{0}}$; using this in
Eq. (\ref{eq3}),
 \begin{equation}
\label{eq4} \frac{GM}{r^{2}} = a \left(\frac{a}{a_{0}}\right)
\end{equation}
which yields,
 \begin{equation}
\label{eq5} a = \left(\frac{GM a_{0}}{r^{2}}\right)^{\frac{1}{2}}
\end{equation}
Also, the equation that relates to the centripetal-acceleration `a' of a star orbiting in a circular
orbit of radius `r' with a velocity `v' in the galaxy is
 \begin{equation}
\label{eq6} a = \frac{v^{2}}{r}.
\end{equation}
Eqs. (\ref{eq5}) and (\ref{eq6}) lead to
\begin{equation}
\label{eq7} v = (GMa_{0})^{\frac{1}{4}}
\end{equation}
Consequently, the velocity of star on circular orbit from the galaxy-center is constant and does not
depend on the distance r; the rotational-curve is `flat'. The relationship $v = (GMa_{0})^{\frac{1}{4}}$
between the flat rotational-velocity v to the observed mass M of the galaxy matches with observed flat
velocity v to luminosity L (known as Tully-Fisher relation). \\\\
It may be noted that the critical acceleration, requiring
MOND-correction, $a_{0} \approx 10^{-10} m/s^{2}$ is negligibly
small a value, that is why we never felt a need for modification in
kinematical-acceleration as mostly on earth the accelerations `a'
are much higher than the critical value `$a_{0}$', hence
$\mu(\frac{a}{a_{0}}) = 1$, thus $a_{effective} = a$. The effect of
MOND is only noticeable to the centripetal-acceleration of galaxy's
rotational-arms wherein the acceleration $a \ll a_{0}$. MOND theory
has been quite successful [19-21], in explaining the
galaxy-cluster-rotation and cosmology-behavior, specially the `flat'
curve of galaxy-arm's rotational-velocity, without any need of
otherwise-necessary dark-matter. The only drawback of MOND is that,
it does not seem to have a good theoretical backing. Though a few
$\mu$-functions have been proposed in literature \cite{ref19}, e.g.,
one such is as following; but the theoretical-backing still lacks
unless some theoretical-support (such as suggested in the present
paper) is provided.
\begin{equation}
\label{eq8} \mu\left(\frac{a}{a_{0}}\right) = \frac{\frac{a}{a_{0}}}{\Big[1 +
\left(\frac{a}{a_{0}}\right)^{2}\Big]^{\frac{1}{2}}},
\end{equation}
which turn the function
\begin{eqnarray}
 \mu\left(\frac{a}{a_{0}}\right) &=& 1 \qquad \textrm {for $a \gg a_{0}$},  \nonumber \\
&=& \frac{a}{a_{0}} \qquad \textrm {for $a \ll a_{0}$}.
\end{eqnarray}
It is not that no work is done on theoretical aspects of MOND, even
Milgrom \cite{ref22} himself has talked about it and discussed
at-length therein \cite{ref22}. But the common sentiment (that still
persists) is expressed occasionally that MOND (successful it may be)
is only a `hypothesis' that `saves the phenomena', and that one day
the origin of MOND-phenomenology may be found. Our present
theoretical-attempt is with this sprit.
\section{A Simple Theoretical Variant of MOND}
The simple key for explaining the low-acceleration-limit MOND (which eliminates the need of dark-matter)
lies in the dark-energy (which is responsible for the acceleration of the expanding universe, however,
the acceleration is extremely-small as estimated in section 4.1).
\subsection{Acceleration due to dark energy}
Initially it was thought that the universe would be decelerating due
to gravity inside, but now it is well established from several clues
[4-12] such as Supernovae observation that the universe is actually
accelerating due to repulsive-gravity of dark-energy. The
deceleration-parameter $q$ is defined as follows in Eq.
(\ref{eq10}). (Note that even though universe is actually
accelerating, but the old-name deceleration-parameter retained; but
$q$ comes out to be actually negative, implying that $\ddot{S}$  is
positive i.e., universe accelerating). Note that though Scale-factor
$S$ (ratio of co-moving distance at previous-time at $Z > 0$ to the co-moving 
distance at present-time $Z = 0$) is dimensionless whereas Size-of-universe (or co-moving-distance)
has dimension of length; but {\it {sometimes}} all these are denoted by the same
symbol S, mainly in view that the scale-factor is proportional to
the size-of-universe (co-moving-distance) and incidentally both the
words begin with the letter S. Sometimes, as in reference \cite{ref2}, 
scale-factor and universe-size are denoted by symbol `a' (but we can not use 
such symbol here because in this paper `a' is used for acceleration). So it is 
better for clarity, if all these are denoted by different symbols as follows; 
scale-factor as S, co-moving distance between two points as D, universe size as 
$D_{max}$, all being function of time due to 
expansion of universe. Thus scale factor $S(t) = \frac{D(t)}{D_{0}}$ or simply 
$S = \frac{D}{D_{0}}$ where D is the co-moving distance in the past (at $Z > 0$) and 
$D_{0}$ is the co-moving distance at present (at $ Z = 0$). Hence it is obvious that 
S is proportional to D. Universe-size is $D_{max}$ i.e., the distance of the visible 
universe-horizon, such that as per Hubble's law $V_{max} = c = H D_{max}$. (Note 
that even if universe-tip may be moving with speed higher than light-speed, as it 
was during inflation, the observable `visible' horizon will be limited by the equation 
$c = H D_{max}$. But in Eq. (\ref{eq11}) what should we use for S ? It seems for galaxy
observations, more appropriately S should be (being proportional) the co-moving-distance D, say,
between the observed galaxy (say, Andromeda galaxy) from the earth
(situated in the Milky-way galaxy). \\\\
The deceleration parameter $q$ is defined by
\begin{equation}
\label{eq10} q = - \frac{S \ddot{S}}{\dot{S}^{2}}.
\end{equation}
Putting the experimental (See the Ref. \cite{ref23}) value of $q = -
0.67$, the expression for the acceleration ($a_{d}$) of the universe
due to dark-energy is given by,
\begin{equation}
\label{eq11} a_{d} = \ddot{S} =  0.67 \times \left(\frac{\dot{S}^{2}}{S}\right).
\end{equation}
Note that from Hubble's law of the expansion, velocity is proportional to 
distance i.e., $V = H D$, H being Hubble's constant. Hubble's law 
$V = H D$ is re-written as $\dot{D} = H D$. This also gives the acceleration 
$\ddot{D} = H \dot{D} = (\frac{\dot{D}}{D})\dot{D} = \frac{\dot{D}^{2}}{D}$ which 
is (almost) exactly of the same-form (equivalent) as the Eq. (\ref{eq11}). Thus, this also 
reinforces the understanding that the co-moving distance D is proportional to 
scale-factor or vice-verse. This also indicates an important possibility that the 
Hubble's expansion is due to dark-energy. Therefore, from (\ref{eq11}), we
derive, by replacing S by D and $\dot{S}$ by $\dot{D}$\\
\begin{eqnarray}
a_{d} &=& 0.67 \times \frac{\dot{D}^{2}}{D} = 0.67 \times \frac{H^{2}D^{2}}{D},  \nonumber \\
&=& 0.67 \times H^{2}D =  0.67 \times H D H,     \nonumber \\
&=& 0.67 \times H\left(\frac{D}{D_{max}}\right)D_{max}H = 0.67 \times H\left(\frac{D}{D_{max}}\right) c,  
\nonumber \\
&=& 0.67 \times c H \left(\frac{D}{D_{max}}\right),
\end{eqnarray}
since it is known from Hubble's law that maximum velocity of our
visible universe-`horizon' can not exceed velocity of light 
(except during the inflation time), hence in limiting case $H \times D_{max} = c$, 
where c is the speed of light. \\\\
Briefly,
\begin{eqnarray}
a_{d} = K \; a_{0} &\approx& 0.67 \times 6 \;\beta \; a_{0} \qquad \textrm {($a_{0} \approx c H
\approx 1.2 \times 10^{-10} m/s^{2}$)}.
\end{eqnarray} 
It is better to express (as Eq. 13 than Eq. 12), for generality \& accuracy, 
$a_{d} = K \; a_{0}$; where $K = 0.67 \times 6 \;\beta $ wherein the factor 6
comes-in because the value of $a_{0} = 1.2 \times 10^{-10} \approx
cH$ as suggested by MOND-proponent M. Milogram, but the actual value
of $cH = 6.8 \times 10^{-10}$ which is about 6 times higher than $a_{0}$.
If  the symbol D is taken as co-moving-distance of a galaxy, say for
example, at 14 million light-years away which is thousand times less than
the visible universe-size $D_{max}$ (14 billion light-years); the distance
ratio-factor $\beta = \frac{D}{D_{max}} = \frac{1}{1000}$. Hence, in that case of
consideration, the value of K could be as $K = \frac{0.67 \times 6}{1000} = 0.004$. 
However, meanwhile, for qualitative theoretical explanation of the new variant of MOND, 
Eq. (13) would be used for subsequent analysis in the present paper.
\subsection{The net effective gravitational (Newtonian) acceleration $a_{n}$ }
Since the universe's expansion is already accelerating with an
acceleration of $a_{d}$; the net effective `kinetic' gravitational
(Newtonian) acceleration `$a_{n}$' due to attractive force $\left(F
= \frac{GMm}{r^{2}}\right)$, will be equal to the modified
`effective kinematical-acceleration' which now equals to the central
centripetal acceleration `a {\it minus} the universe/galaxy's slow
acceleration $a_{d}$', i.e., $a_{effective} = a - a_{d}$. For
equilibrium,
\begin{eqnarray}
a_{n}  &=& a_{effective} = a - a_{d},  \nonumber \\
&=& a - K \; a_{0}, \nonumber \\
&=& a\left(1 - K \frac{a_{0}}{a}\right).
\end{eqnarray}
The theoretical $\mu$-function obtained here is simply $\left(1 -
K \frac{a_{0}}{a}\right)$, which becomes 1 for normal and large
value of a, and is less than 1 if the value of a is extremely-small
$(a \ll a_{0})$.
\begin{figure}[ht]
\centering
\includegraphics[width=11cm,height=10cm,angle=0]{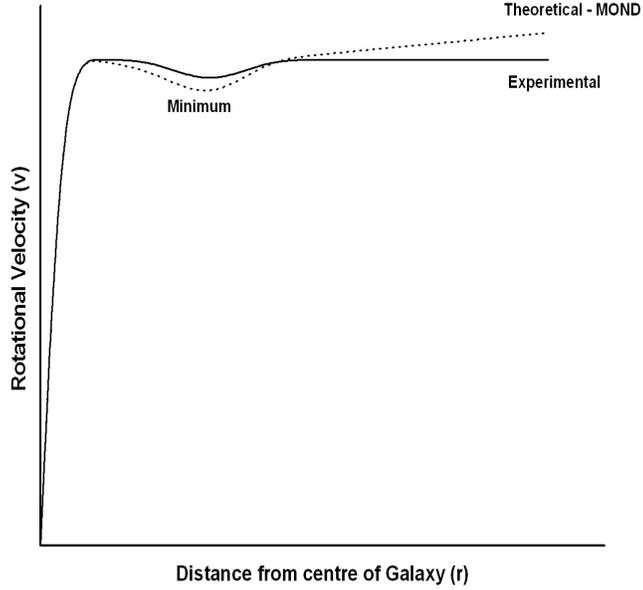}
\caption{Rotational velocity versus distance from Galaxy-centre:
experimental `flat' curve with a Dip and a more-or-less flat
theoretical curve with a minimum}
\end{figure}
\begin{figure}[ht]
\centering
\includegraphics[width=8.cm,height=12.0cm,angle=0]{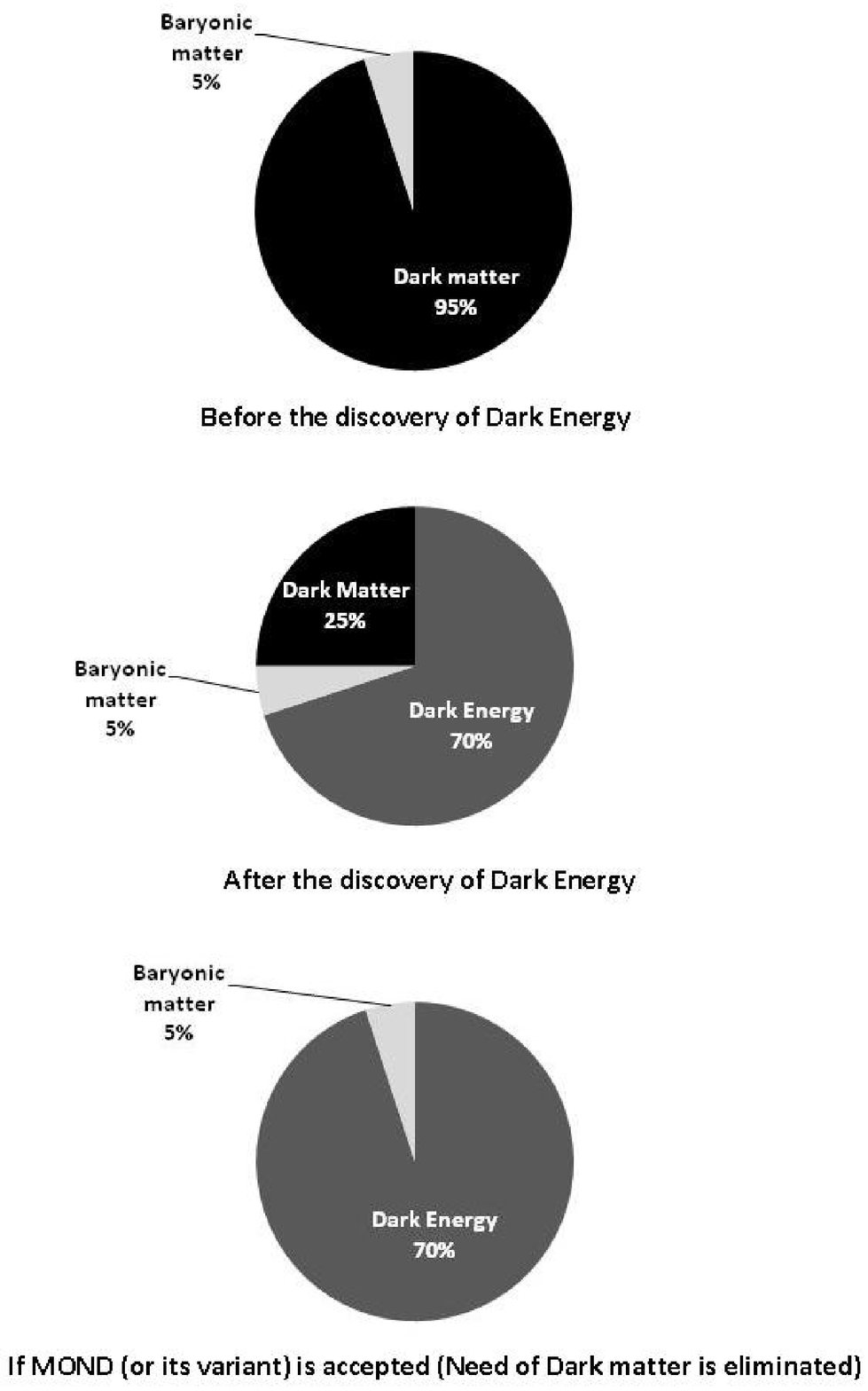}
\caption{Possible Dark-constituents slices of the Universe}
\end{figure}
\subsection{Galaxy's rotational velocity estimate}
Since gravitational (Newtonian) acceleration $a_{n} = \frac{GM}{r^{2}}$ and that centripetal acceleration
toward centre is known as $a = \frac{v^2}{r}$, the previous equation (14) reduces to
\begin{equation}
\label{eq15} v = \left(\frac{GM}{r} + K\; a_{0}\; r\right)^{\frac{1}{2}}.
\end{equation}
For minimum  $v\;( or ~ v^{2})$, $\frac{dv^2}{dr} = 0$, gives that minimum velocity will occur at  a
certain value $r^{*}$ as follows; this explains why there is a little dip in the generally-considered
`flat' rotational-velocity curve.
\begin{equation}
\label{eq16} r^{*} = \left(\frac{GM}{K\; a_{0}}\right)^{\frac{1}{2}}.
\end{equation}
The success of the present approach lies in the fact that not only
the Eq. (\ref{eq15}) gives more or less a `flat' curve (due to
decrease in $v$ due to the first term $\frac{GM}{r}$ and increase in v
due to the second term $K\; a_{0}\; r$), but also predicts a dip
(minimum) of this so-called `flat' curve (Fig.3); note that these
two terms together has a square-root sign (1/2 power) over it,
trying to make the curve flatter after the dip. The explanation and
agreement is good qualitatively, but detailed quantitative analysis
is needed  to be done.

\section{Re-slicing of the Constituent(s) of the Universe}
Before the discovery of dark-energy, till 1998, it was thought that
the dark-constituent (as shown in Fig.4 by shaded area) was having a
single big slice of $95\%$ dark-matter with at the most a $5\%$ tiny
slice of baryonic matter (stars, planets, galaxies, black-hole,
white dwarf etc.). But after the discovery of dark-energy; the
bigger dark-constituent slice was subdivided into two dark slices:
smaller dark slice (about) $25\%$ of dark-matter and the bigger dark
slice (about) $70\%$ of dark-energy. However, if MOND model which
works well and eliminates the need of of dark-matter all-together,
is accepted; then the composition of the dark-constituent(s) will be
different, all the $95\%$ would be dark-energy, as no dark-matter needed 
with MOND.\\\\ 

Both the dark matter and the dark energy are suggestions. But since, the 
acceleration (though small) of universe is now well established, the dark-energy seems more
realistic than before. Dark-matter is rather quite strange and
wizard; several schools of thoughts (cold and hot dark-matter and
its candidates) have been proposed with no success as yet, as if the
dark-matter is nothing but castle-in-air. In this situation; if MOND
(which is an alternative to dark-matter theory) is accepted
(specially after having a theoretical backing), the need of
dark-matter is eliminated completely, making the dark slice of
dark-energy alone. This may however, require a little modification
(fine tuning) of  the dark-energy theory.
\section{Conclusions}
Though there exists a successful empirical MOND theory, as proposed by the Israel-scientist M. Milgrom, for
explaining the anomalous rotational-velocity of galaxies and beyond, but its drawback is that it lacks
theoretical backing. In the present paper a novel variant of MOND is proposed on theoretical footing; this too
at-present is able to explain the rotational-velocity curve qualitatively fairly well. MOND (or its variant)
eliminates the need of dark-matter, the theoretical basis for the MOND-variant proposed herein lies in the
acceleration of universe caused by dark-energy; meaning-by that it is the presence of dark-energy that
eliminates the need of dark-matter. Though the present theory also predicts a dip in the so-called `flat'
curve in-agreement with the experimental-curve, more work on quantitative level needs to be done to establish
the proposed-theory firmly. The present work gives a clear theoretical understanding of the so-far empirical
MOND and starts a new beginning in this direction. The secret of high velocity depicted in flat
rotational-velocity curve of big galaxy lies in the universe's small acceleration (Eq. 13), caused by
the mighty dark-energy possibly having its genesis \cite{ref2} in the tiny nucleus.
\section*{Acknowledgements}
The authors are thankful to Sushant Gupta for critical study of this research-paper. One of the authors 
(R. C. Gupta) is thankful to Project-students (S. Goyal, A. Kumar, I. Ansari \& R. Singh) of IET
Lucknow, and GLA, Mathura; and the other author (A. Pradhan) is thankful to IMSc., Chennai (Madras) 
for providing facility under associateship scheme; the places where the research-work is
initiated and completed.  


\noindent
\begin{thebibliography}{99}
\bibitem {ref1}
Beiser, A., Concept of Modern Physics, 6th Ed., Mc Graw Hill, New York (2003).
\bibitem {ref2}
Gupta, R. C. and Pradhan, A., Int. J. Theor. Phys. {\bf 49}, 821 (2010).
\bibitem {ref3}
Kaku, M., Beyond Einstein, Anchor Books, New York (1995).
\bibitem {ref4}
Perlmutter, S., et al., Nature {\bf 391}, 51 (1998).
\bibitem {ref5}
Perlmutter, S., et al., Astrophysics J. {\bf 517}, 565 (1999).
\bibitem {ref6}
Riess, A. G., et al., Astron. J. {\bf 116}, 1009 (1998).
\bibitem {ref7}
Riess, A. G., et al., Astron. J. {\bf 607}, 665 (2004).
\bibitem {ref8}
Tonry, J. L., et al., Astrophys. J. {\bf 594}, 1 (2004).
\bibitem {ref9}
Garnavich, P., et al., Astrophys. J. Lett. {\bf 493}, L53 (1998).
\bibitem {ref10}
Schmidt, B. P., et al., Astrophys. J. {\bf 507}, 46 (1998).
\bibitem {ref11}
Bahcall, N. A., et al., Science {\bf 284}, 1481 (1991).
\bibitem {ref12}
Spergel, D.N., et al., Astrophys. J. Suppl. {\bf 148}, 175 (2003).
\bibitem {ref13}
Zwicky, F., Helvetica Physica Acta {\bf 6}, 110 (1933).
\bibitem {ref14}
Zwicky, F., Astrophys. J. {\bf 86}, 217 (1937).
\bibitem {ref15}
Rubin, V. and Ford, W. K. Jr, Astrophys. {\bf 159}, 379 (1970).
\bibitem {ref16}
Rubin, V., Thonnard, N. and Ford, W. K. Jr, Astrophys. J. {\bf 238}, 471 (1980).
\bibitem {ref17}
Milgrom, M., Astrophys. J. {\bf 270}, 365 (1983).
\bibitem {ref18}
Milgrom, M., Astrophys. J. {\bf 270}, 371 (1983).
\bibitem {ref19}
Scarpa, R., arXiv:0601478 (2006).
\bibitem {ref20}
Sanders, R. H., astro-ph/9710335 (1997).
\bibitem {ref21}
Bekenstein, J. D., astro-ph/0701848 (2007).
\bibitem {ref22}
Milgrom, M., J. New Astronomy Reviews {\bf 46}, 741 (2002), (arXiv: astro-ph/0207231v2).
\bibitem {ref23}
Freedman, W. L. and Turner, S. S., Rev. Mod. Phys. {\bf 75}, 1433 (2003).
\end{thebibliography}
\end{document}